\documentclass[referee]{raa}            % referee version: for submission

\usepackage{graphicx,times}             %for PS/EPS graphics inclusion, new
\usepackage{natbib}
\usepackage{amssymb,amsmath}
\usepackage{threeparttable}
\usepackage{adjustbox}
\usepackage{tabularx}
\usepackage{amsmath}
\bibpunct{(}{)}{;}{a}{}{,}
\usepackage{caption}

\usepackage[pagebackref=true]{hyperref}
\begin{document}

  \title{A Search for Exoplanets in Open Clusters and Young Associations based on TESS Objects of Interest
}
%   \subtitle{I. Place Your Subtitle Here}

   \volnopage{Vol.0 (20xx) No.0, 000--000}      %%preserved for Editor. Don't remove!
   \setcounter{page}{1}          %%starting page, preserved for Editor. DOn't remove!

   \author{Qinghui Sun\inst{1}
   \and Sharon Xuesong Wang\inst{1}
    \and Tianjun Gan\inst{1}
   \and Andrew W. Mann\inst{2}
   }

    \institute{Department of Astronomy, Tsinghua University, Beijing 100084, People’s Republic of China; {\it qingsun@mail.tsinghua.edu.cn; sharonw@mail.tsinghua.edu.cn} \\
    \and
   	{Department of Physics and Astronomy, The University of North Carolina at Chapel Hill, Chapel Hill, NC 27599, USA} \\
\vs\no
   {\small Received 20xx month day; accepted 20xx month day}}

\abstract{We report the results of our search of planet candidates in Open Clusters and Young Stellar Associations based on the TESS Objects of Interest Catalog. We find one confirmed planet, one promising candidate, one brown dwarf, and three unverified planet candidates in a sample of 1229 Open Clusters from the second Gaia data release. We discuss individual planet-star systems based on their basic parameters, membership probability, and the observation notes from the ExoFOP-TESS website. We also find ten planet candidates (P $>$ 95\%) in Young Stellar Associations by using the BANYAN $\Sigma$ Bayesian Algorithm. Among the ten candidates, five are known planet systems. We estimate the rotation periods of the host stars using the TESS light curves and estimate their ages based on gyrochronology. Two candidates with periodic variations are likely to be young planets, but their exact memberships to Young Stellar Associations remain unknown.
}

   \authorrunning{Sun, Q. et al.}            %author_head in even pages
   \titlerunning{Exoplanets in Open Clusters and Young Associations }  % title_head in odd pages

   \maketitle
%% The author head (on even pages) and the title head (on odd pages) will be
%% automatically extracted from \author{} and \title{}. Whenever the title is too long,
%% you will be asked to supply a shorter one by inserting either \authorrunning{} or
%% \titlerunning{} before \maketitle. Anyway, you can specify your own heads.
%%
%%
%% Note: In the following text body of your manuscript, please note several differences from
%%       other major journals:
%% (1) \subsection{Please Capitalize the First Letter of Each Notional Word in Subsection Title}
%% (2) Please Capitalize the First Letter of Each Notional Word in all tables' captions

%
%________________________________________________ sections below
%
\section{Introduction}           %% first-level sections will be auto-capitalized
\label{sect:intro}

Young planets ($<$ 100 Myr) are especially useful in understanding the early stages of planet formation and evolution when major physical changes happen (\citealt{Pollack96}, \citealt{Lopez13}, \citealt{Owen13}). Planet has the potential to migrate, either by planet-disk interactions (\citealt{Nelson17}, \citealt{Kley12}) or planet-planet interactions (\citealt{Chatterjee08}), and as a result may induce atmospheric interactions (\citealt{Fortney11}; \citealt{Lammer03}). The formation of planets and the characteristics of planet systems (e.g. mass, compositions) are correlated with the chemical abundances of their host stars (\citealt{Ramirez11}, \citealt{Brugamyer11}, \citealt{Saffe17}). These physical processes could be explored by studying planet systems of different ages.

Ages are key parameters in understanding the timing of different physical processes, and are strongly correlated with the internal structure and evolutionary processes of these young planets. Coeval populations (e.g. Globular Cluster, Open Cluster, young stellar association) have better determined stellar parameters (e.g. age, chemical composition) and clearer stellar evolution history than individual field stars (\citealt{Sun21}). For example, the age of an Open Cluster could be derived from photometry by fitting an isochrone to the Color-Magnitude Diagram of member stars (\citealt{Sun20}). These coeval populations are thus ideal for performing systematic planet searches and statistical comparisons of planet systems at different ages.

The Transiting Exoplanet Survey Satellite (TESS, \citealt{Ricker15}) identifies exoplanet candidates around nearby bright stars thanks to its full sky coverage, providing a valuable source for young exoplanet candidates. The TESS Object of Interest Catalog (\citealt{Guerrero21}) includes 5038 planetary candidates from the TESS survey, among which only $\sim$ 170 are confirmed planets\footnote{From the TESS publication website: \url{https://tess.mit.edu/publications}, retrieved January 2 2022.}, leaving a large number of candidates waiting to be confirmed. Several searches of planets in Open Clusters have been successful in the nearby Hyades ($\sim$750 Myr, \citealt{Quinn14}) and Praesepe ($\sim$790 Myr, \citealt{Quinn12}, \citealt{Obermeier16}). The TESS Hunt for Young and Maturing Exoplanets (THYME) survey has been focused on finding transiting planets in nearby young stellar associations (\citealt{Mann21,Tofflemire21,Newton21,Mann20,Rizzuto20,Newton19}). But the number of planetary systems in these coeval populations is still low. We expect to find more planet systems in Open Clusters and Young Stellar Associations based on a systematic TESS search.

TESS has reported individual discovery of planets in Open Clusters (e.g. \citealt{Bouma20}). Based on the TESS mission, a few statistical assessment of the frequency of planets and age-planet radius distribution relations in Open Clusters have been reported (\citealt{Nardiello20}; \citealt{Nardiello21}). Different from these statistical studies of planet properties, in this paper we search over 5000 planet candidates from the TESS Object of Interest Catalog in 1229 Open Clusters (\citealt{Cantat18}) from the second Gaia data release (\citealt[hereafter Gaia DR2][]{Gaia18}. We perform the analysis to see if any candidates happen to be an Open Cluster member, discuss individual candidates of interest and promising ones for follow-up studies.

The Open Cluster members are tightly bound together by gravity, with typical ages between several ten Myrs to a few Gyrs. The TESS survey mainly searches for exoplanets around nearby bright stars, so the number of TESS planets detected in Open Clusters may be limited as a majority of Open Clusters are faint and old ($>=$ 100 Myr). Young stellar associations are loosely grouped together young stars with coeval populations, so they have the same advantages as Open Clusters as discussed above. We even expect to find more planets in young associations as they are mostly nearby and bright compared to Open Clusters. In this paper, we evaluate the membership probability of TOI candidates to a list of 27 young associations with ages between 1-800 Myr by using the BANYAN $\Sigma$ Multivariate Bayesian Algorithm (\citealt{Gagne18}). We measure rotation periods of the host stars using TESS light curves and estimate their ages based on gyrochronology. We then discuss individual candidates and promising ones for follow-up studies.

This paper is arranged as follows: Section \ref{sec:data} shows our data analyses procedure, search results in Open Clusters and in young associations; Section \ref{sec:discuss} discusses each individual candidates of interest and promising ones for follow-up studies; Section \ref{sec:summary} summarizes the paper.

\section{Data Analysis} \label{sec:data}

The TESS Object of Interest (TOI) Catalog includes basic parameters of over 5000 exoplanet candidates (e.g. orbital periods, transit depths/size) and main stellar parameters of their host stars (e.g. effective temperature, log g). The TOI catalog is available online \footnote{Data retrieved on January 2 2022 from \url{https://tev.mit.edu/data/collection/193/}}, and also by querying through the Mikulski Archive for Space Telescopes (MAST) list.\footnote{\url{https://astroquery.readthedocs.io/en/latest/mast/mast.html}}

\subsection{Open Cluster}

By using parallax and proper motion information from Gaia DR2, \citet[CG18]{Cantat18} identified and derived membership probabilities for over 0.4 million stars in 1229 Open Clusters. We match TOI candidates with the list of possible Open Cluster members using the coordinates, parallax, and proper motion information. The TESS Follow-up Observing Program (TFOP) working group (TFOPWG\footnote{\url{https://tess.mit.edu/followup/}}) aims to confirm and characterize planet candidates identified by TESS. One of the subgroups of TFOP -- Seeing-limited Photometry (SG1) -- performs ground-based photometric followup observations to better determine the transit ephemeris and identifies false positives (\citealt{Collins18}). The TFOP-SG1 disposition is publicly available from the Exoplanet Follow-up Observing Program for TESS (ExoFOP-TESS) website\footnote{\url{https://exofop.ipac.caltech.edu/tess/}}.

After combining the TOI and TFOF-SG1 disposition, we exclude all the false positives from our search results. The false positives include TOI 1535.01 (nearby eclipsing binary), TOI 861.01 (nearby eclipsing binary), TOI 517.01 (eccentric eclipsing binary), TOI 2959.01 (nearby eclipsing binary), TOI 1918.01 (eclipsing binary), TOI 4397.01 (nearby eclipsing binary), TOI 1188.01 (nearby eclipsing binary), TOI 1321.01 (nearby eclipsing binary), TOI 1497.01 (nearby eclipsing binary), TOI 2451.01 (nearby planet candidate), TOI 496.01 (eclipsing binary), and TOI 580.01 (possible nearby eclipsing binary). We summarize our final list of planet candidates in Open Clusters in Table \ref{tab:OC} (the false positives are not included here). The table shows selected parameters from the TOI catalog, the associated errors and descriptions of each parameters could be found in \citet{Guerrero21}. Note that all the candidates are singles and have ``.01'' in their TOI Id. Column 14  is the candidate's membership probability to the specific Open Cluster in Column 15 retrieved directly from CG18.

\begin{table}
	\resizebox{\textwidth}{!}{
		\centering
	\begin{threeparttable}
		%\begin{center}
		\caption[]{Planet Candidates in Open Clusters.}\label{tab:OC}
		\tiny
		\begin{tabular}{cccccccccccccccccc}
			\hline\noalign{\smallskip}	
			TIC Id$^1$ & TOI Id$^1$ & RA$^1$ & DEC$^1$ & TESS$^1,^2$ & TMag$^1$ &  Orbital$^1$ & Transit$^1$ & Transit$^1$ & $T_{eff}^1$& log g$^1$ & Star$^1$ & Planet$^1$ & P$_{mem}^3$ & OC$^3$ & TFOP-SG1$^4$ \\
			&  & &  & Disposition &  & Period & Duration & Depth & & & Radius & Radius & & & Disposition \\
			&  & degree & degree & & mag & days & hours & ppm & K & & $R_{\odot}$ & $R_E$ & & & \\
			\hline\noalign{\smallskip}
			142938659 & 4668.01 & 52.6471645 & -35.2036955 & PC & 13.837 & 1.0049552	& 0.92 & 8600 & 3210.0 & 4.80594 & 0.42 & 5.19 & 1.0 & Alessi 13 & PC \\
			59859387 & 1881.01 & 110.3697368 & -45.5677871 & PC & 10.276 & 1.120 & 2.387 & 1739 & 6564.4 & 4.26 & 1.47 & 9.90 & 0.7 & Alessi 3 & VPC+ \\
			460205581 & 837.01 & 157.037444	& -64.505257 & CP & 9.916 & 8.325 & 1.647 & 4360 & 6513 & 4.53694 & 1.01 & 7.60 & 0.9 & IC 2602 & VP \\
			443115574 & 2538.01 & 95.226324	& -7.298935 & PC & 10.523 & 2.910 & 3.606 & 1480 & 7340.8 & 4.09 & 3.50 & 12.26 & 0.8 & NGC 2215 & PC \\
			410450228 & 681.01 & 117.89497 & -60.412452 & PC & 10.656 & 15.780 & 3.386 & 6808 & 7447 & 3.79 & 1.69 & 18.27 & 0.4 & NGC 2516 & VPC+ \\
			180987952 & 581.01 & 130.2615216 & -41.4428264& PC & 9.529 & 1.389 & 1.476 & 630 & 10585 & 4.35 & 1.84 & 4.84 & 0.3 & Trumpler 10 & PPC \\
			\noalign{\smallskip}\hline
		\end{tabular}
		\begin{tablenotes}
			\item[1] TESS Input Catalog (TIC) Id and key parameters, data retrieved from the TOI Catalog (\url{https://tev.mit.edu/data/}), associated errors and descriptions of each parameters could be found in \citet{Guerrero21}.
			\item[2] ``PC" stands for ``possible candidate", ``CP" stands for ``confirmed planets" (Master disposition keywords from the TOI catalog).
			\item[3] The membership probability of the planet-hosting star to an existing Open Cluster.
			\item[4] Master Dispositions from TFOP-SG1. ``VPC+" stands for ``Verified Achromatic Planet Candidate", ``VP" stands for`` validated planet", ``PC" stands for ``Planet Candidate",  ``PPC" stands for "Promising Planet Candidate" (disposition keywords can be found on the ExoFop website).
		\end{tablenotes}
		%\end{center}
	\end{threeparttable}
}
\end{table}

We find one confirmed planet, one brown dwarf, and four planet candidates in Open Clusters. The confirmed planet is TOI 837.01, discovered by \citet{Bouma20} as a transiting planet in the open cluster IC 2602. TOI 681.01, the transiting companion of TOI 681, is a confirmed Brown Dwarf (\citealt{Grieves21}). We remind the reader that TOI catalog mainly includes bright, nearby planets, whereas many Open Clusters are older, relatively far and faint. The current TOI catalog also includes the ones from the Quick-Look Pipeline (QLP) faint star search, which extends the magnitude limit to T $\sim$ 14.0 mag (TESS magnitude, \citealt{Kunimoto21}), our search results are likely constrained by the magnitude limit. Future searches of fainter stars in a broader region may lead to more discoveries of planet systems in Open Clusters.

\subsection{Young Association}

\citet{Gagne18} developed a BANYAN $\Sigma$ Bayesian Algorithm with the multivariate Gaussian functions in six-dimensional space to calculate the membership probability of a star belonging to young associations. The algorithm includes bona fide members of 27 young associations within 150 pc of the Sun, with typical ages between 1 - 800 Myr. 

We use the algorithm to identify planet candidates with previously unknown membership in young associations for all TOIs. We incorporate the TOI and the publically-available TFOP-SG1 disposition to eliminate false dectections (mainly eclipsing binaries). The false detections include TOI 447.01 (eclipsing binary), TOI 450.01 (ambiguous planetary candidate), TOI 456.01 (likely nearby eclipsing binary), TOI 4636.01 (eclipsing binary), TOI 235.01 (likely eclipsing binary from the TFOPWG notes), TOI 831.01 (ambiguous planetary candidate), TOI 1047.01 (nearby planet candidate), TOI 278.01 (likely stellar variation), TOI 2496.01 (likely eclipsing binary), TOI 935.01 (eclipsing binary), TOI 734.01 (likely eclipsing binary), TOI 1433.01 (ambiguous planetary candidate), TOI 919.01 (eclipsing binary), TOI 1500.01 (nearby eclipsing binary), and TOI 224.01 (ambiguous planetary candidate).

We show the search result in Table \ref{tab:YA}, which only includes planet candidate (PC) and known planet (KP). The table shows basic parameters of the planet candidates and their membership probability to young stellar associations.

\begin{table}
	\begin{adjustbox}{angle=90}
	\resizebox{1.5\textwidth}{!}{
		\centering
	\begin{threeparttable}
		%\begin{center}
		\caption[]{Probable Planet Candidates in Young Associations.}\label{tab:YA}
		\tiny
		\begin{tabular}{cccccccccccccccccc}
			\hline\noalign{\smallskip}	
			TIC Id & TOI Id & RA$^1$ & DEC$^1$ & TOI & TMag$^1$ & Orbital$^1$ & Transit$^1$ & Transit$^1$ & $T_{eff}^1$ & log g$^1$ & Star$^1$ & Planet$^1$ & P(YA)$^2$ & LIST\_PROB\_YA$^3$ & SG1$^4$ & Notes \\
			&  & & & Disposition &  & Period & Duration & Depth & & & Radius & Radius & & & Disposition & \\
			&  & degrees & degrees &  & mag & day & hour & ppm & K & & $R_{\odot}$ & $R_E$ & & & Disposition & \\
			\hline\noalign{\smallskip}
			410214986 & 200.01 & 354.914556	& -69.195787& KP & 7.771 & 8.14 & 3.32 & 3576 & 5414 & 4.4 & 0.95 & 6.95 & 0.999 & Tucana-Horologium (THA) & VP & DS Tuc Ab (THYME planet, \citealt{Newton19})\\
			\hline
			142938659 & 4668.01 & 52.6471645 & -35.2036955 & PC & 13.837 & 1.00 & 0.92 &8600 & 3210 & 4.81 & 0.42 & 5.19 & 0.998 & $X^1$ For (XFOR) & PC & Unlikely a YA member \\ 
			\hline
			359357695 & 1880.01 & 182.7268485 & -75.1318983 &PC& 13.061 & 1.73 & 1.73 &9681 &3631&4.69&	0.56&6.57&0.998& $\epsilon$ Chamaeleontis (EPSC) &VPC& Age discrepancy with EPSC, likely wrong YA disposition \\
			\hline
			441420236 & 2221.01 & 311.2897 & -31.3409 &KP& 6.755 & & 3.57 & 2872	&3588& 4.60 &0.70&4.03&	0.994 & $\beta$ Pictoris ($\beta$ PMG) &P	& AU Mic b (\citealt{Plavchan20}) \\
			\hline
			34077285.01 &880.01& & &PC& &6.39 & 2.61 & 3510 & & & & 5.04 & & &VPC & Unlikely a YA member \\
			34077285.02 &880.02& 94.1644691 & -13.987437&PC&9.256&2.57&2.48&740&4935&4.52&0.82 & 2.78 &0.99 & Argus (ARG) &VPC & potential multi-planet system \\
			34077285.03 &880.03& & &PC& & 14.33 & 1.60 & 990 & & & 0.82 & 2.87 & & & & \\
			\hline
			383390264&1098.01 & 192.592205 & -88.121055&PC&8.739&10.18&2.85	&715&6153&4.35&1.20& 3.26 &0.990& Octans (OCT)$^5$ &P&  HD 110082 (THYME planet, \citealt{Tofflemire21}) \\
			\hline
			360156606&1227.01 & 186.76803 & -72.45179&PC&13.757	&27.36&	4.14&27234&3050&4.44	& 1.00 &16.52&	0.986 & $\epsilon$ Chamaeleontis (EPSC)$^6$ &VPC	& TOI 1227 b (THYME planet, \citealt{Mann21})\\
			\hline
			67646988 & 1779.01 & 147.7690653 & 35.96929602&KP&12.437&1.88&	1.24 & 102854 & 3138 & 4.92 & 0.31 & 9.93 & 0.968 &AB Doradus (ABDMG) & KP & LP 261-75 b (\citealt{RW06}) \\
			\hline	
			142937186&2427.01 & 52.2909729 &-31.3630447 &PC&9.014&1.31&0.63&560&	4072&4.58&0.68&2.00& 0.954 & Argus (ARG, 73); Carina-Near (CARN, 27) & VPC & Unlikely a YA member \\
			\hline
			150151262.01 &712.01 & & & & &9.53&1.83	&959& & & & 2.52 & & & & Age discrepancy with ABDMG, likely wrong YA disposition\\
			150151262.02 &712.02 & 92.936152 & -65.825972&PC&9.845&51.70&4.76	&1882&4504&4.59&0.72&3.14&0.953 & AB Doradus (ABDMG) & VPC+ & potential multi-planet system\\
			150151262.03 &712.03 & & & & & 53.75 &5.59	&1558& & &	& 3.26 & & & & \\
			150151262.04 &712.04 & & & & &679&5.56	&1557& & &	&2.789 & & & & \\
			\hline
			89256802 & 457.01 & 58.7249	& -26.4237&PC	&13.801	&1.18&	0.99	&14881&3054&5.09&0.51&7.72&0.946& Argus (ARG) &VPC-& \\
			\hline
			20318757.01 & 1027.01 & 167.1337084 & -29.6531909& PC	& 10.21	&3.28	&1.38&	1220&4272&4.6&0.68&2.70&0.945& Argus (ARG) &	VPC+& potential multi-planet system \\
			20318757.02 & 1027.02 & & &	&	&11.03	&3.16& 1770& & & &3.12& & &	 & \\
			\hline
			157115010 &	4319.01	& 100.9720399&-43.89332391& PC &10.82& & 7.73 & 6743 & 4438 & 4.44 & 0.73	&6.55&	0.940&	Argus (ARG)& STPC & \\
			\hline
			464646604&4399.01 & 287.7410437 & -60.272202&PC&7.758&7.71&2.43 &728 & 5985&4.4 &1.09& 3.00 &0.890 & AB Doradus (ABDMG) &PC	& \\
			\hline
			290348383&1099.01 & 328.713617 & -77.338017&PC&7.394&6.44 &	1.83&827&4867&4.44&0.80&2.46	&0.862& Carina-Near (CARN) &VPC& \\
			\hline
			260647166.01&1233.01& & & CP & &14.18&3.85 & 846 & &4.44& &2.63& & & & HD 108236 d (\citealt{Daylan21}) \\
			260647166.02&1233.02& 186.574587 & -51.362819&CP&8.618&19.59&4.17 & 1161 &5724&4.44& 0.86 & 3.22 &0.846 & AB Doradus (ABDMG) &VP& HD 108236 e \\
			260647166.03&1233.03& & & CP & &6.20&3.00 & 698 & & & &2.46& & & & HD 108236 c \\
			260647166.04&1233.04& & & CP & & 3.80 & 2.31 & 322 & & & &1.67& & & & HD 108236 b \\
			\hline
			406672232&1263.01 & 309.353389 & 22.65461489 &PC&8.533	&1.02&	1.06 &258 &5098& 4.55 &	0.82 &1.43 &0.762 & AB Doradus (ABDMG) &PC& \\
			\hline
			360630575 & 1097.01 & 189.7766743 & -74.5739909& PC & 8.722 & 9.19 & 2.48 & 420 &	5876 & 4.48 & 0.98 & 2.17 & 0.702 & Lower Centaurus Crux (LCC) & CPC & \\
			\hline
			429302040 &1905.01 & 188.386857	& -10.146147&KP	&10.474	&5.72 &	2.65	&25549	&4233	&4.64	&0.73	&12.84	&0.681	& Argus (ARG)	&KP	& WASP-107 b (\citealt{Anderson17}) \\
			\hline
			77253676 &697.01 & 69.704057 & -36.681466&PC	&9.304	&8.61&	3.68	&555	&5447	&4.40 &	0.97	&2.62	&0.659 & Argus (ARG) & CPC & \\
			\hline
			391903064.01 &3353.01 & 106.529799 & -75.819712 & PC & 8.751 & 4.67 &3.04	&647	&6365	&4.55	&1.02&	2.33 &0.638	&Lower Centaurus Crux (LCC, 69); AB Doradus (ABDMG, 31)	&PC& \\
			391903064.02 &3353.02 & & & & & 8.82 &3.11	&490& &	& &	2.33 & & & & \\
			\hline
			148883384 &2522.01 & 107.1185305 & -12.1478068&PC	&9.124	&2.10 &	4.37	&350	&5531	&4.45	&0.97	&1.67	&0.631	& Columba (COL) &PC	& \\
			\hline
			451645081 &783.01 & 172.15656 & -54.784245&PC &9.859	& 16.22 &2.10 & 1262 &4499&4.5&0.67	&2.394&0.538& Argus (ARG) & CPC?& \\
			\noalign{\smallskip}\hline
		\end{tabular}
		\begin{tablenotes}
			\item[1] Selected parameters from the TOI Catalog. Due to space constraints, errors associated with these parameters are not shown in the table, but can be found in the online catalog. Detailed descriptions of each parameters can be found in \citet{Guerrero21}.
			\item[2] Probability that the planet-hosting star belongs to a young association, calculated from BANYAN $\Sigma$ Bayesian Algorithm.
			\item[3] A list of young associations that the planet-hosting star probably belongs to. When two associations are given, individual probability (in percentage) is included in the bracket.
			\item[4] TFOP-SG1 Disposition, abbreviations have the same meaning as in Table \ref{tab:OC}. In addition to those already appeared in Table \ref{tab:OC}, ``VPC" stands for ``Verified Planet Candidate"; ``P'' stands for ``confirmed planet"; ``KP" stands for ``Known planet" (disposition keywords can be found on the ExoFop website).
			\item[5] HD 110082 is a member of the newly discovered 250 Myr Association MELANGE-1, rather than Octans association from the BANYAN $\Sigma$ algorithm.
			\item[6] TOI 1227 is a member of the Lower Centaurus Crux (LCC) OB association (\citealt{Mann21}), rather than $\epsilon$ Chamaeleontis (EPSC) from the BANYAN $\Sigma$ algorithm.
		\end{tablenotes}
		%\end{center}
	\end{threeparttable}
}
\end{adjustbox}
\end{table}

\section{Discussions} \label{sec:discuss}

In this section, we discuss planet candidates listed in Table \ref{tab:OC} and Table \ref{tab:YA}, some of which may be interesting for follow-up studies. We recommend the readers to check ExoFop-TESS for public comments, available data products, and contact the TFOP group for latest updates on candidate status.

\subsection{Open Clusters}

\begin{figure*}
	\centering
	\includegraphics[width=0.9\textwidth, trim={8cm 0 8cm 0}]{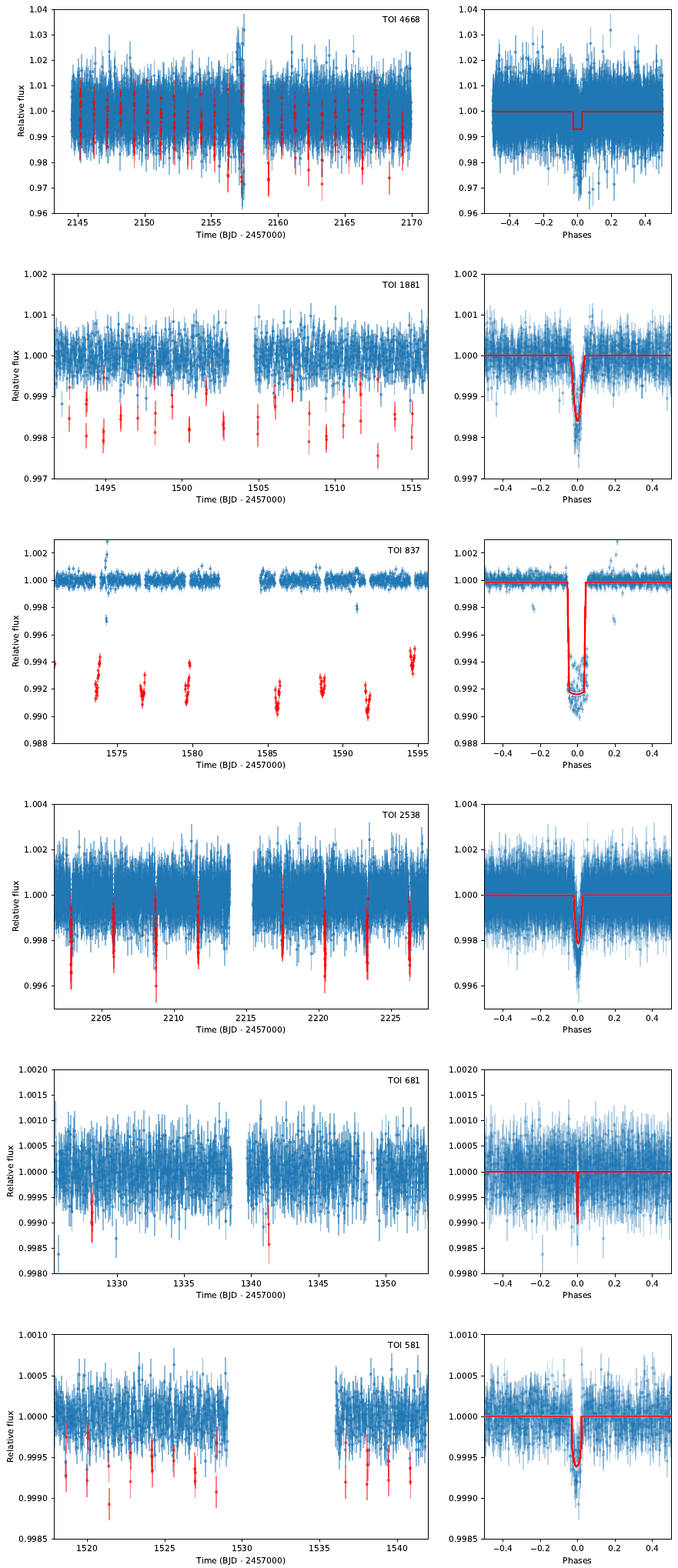}
	\caption{TESS detrended light curves (left panel) and phase-folded transit light curves (right panel) for planet candidates in OC. From top to bottom are TOI 4668, TOI 1881, TOI  837, TOI 2538, TOI 681, and TOI 581. Potential transits are marked using red dots in the left panel, and the fitting results are plotted using red lines in the right panel.}
	\label{fig:transit_OC}
\end{figure*}

To give readers an overall impression of the planet candidates listed in Table \ref{tab:OC}, we detrend, remove outliers, and normalize the flux of the TESS light curves with the \texttt{lightkurve} package in Python, and then further remove stellar variability to create phase-folded light curves by using the \texttt{Juliet} package in Python. The TESS light curves and phase-folded transits are shown in Figure \ref{fig:transit_OC} for candidates in Table \ref{tab:OC}. The flux used to plot the figure is pre-search Data Conditioning SAP flux (PDCSAP, systematic trends removed and light dilution corrected. ). TOI 4668 does not have PDCSAP available, so we use the QLP detrended light curves instead (KSPSAP flux). Based on the transit light curves, public comments, and publically available data products, in the rest of this section we discuss individual candidates listed in Table \ref{tab:OC}.

TOI 4668 is an early M dwarf found by the QLP faint star search, its transit signal is V-shaped and needs further validation.

TOI 1881 is a hot F star reported as a Community TESS Objects of Interest (CTOI). The latest TFOP-SG1 note says the object is an achromatic verified planet candidate (VPC+). Precise radial velocity measurement may be tough due to its fast rotating. Further observations of the Rossiter-McLaughlin effect (\citealt{Triaud18}) could measure the spin-orbit angle and help validate the planet. 

TOI 837.01 is a confirmed young transiting planet in the Open Cluster IC 2602 (\citealt{Bouma20}).

TOI 2538 is a member of NGC 2215, whose spectrum looks very broad ($V_{ROT}$ = 180 km s$^{-1}$) with prominent $H_{\alpha}$ line. TFOP-SG1 note says the object is located in a crowded field and the TESS team reports centroid shift. The target is hard to validate or confirm.

TOI 681.01 is a confirmed brown dwarf (\citealt{Grieves21}).

TOI 581 (A0 or B star) is less likely to be a member of Trumpler 10, speckle images (HRCam) and spectra (CORALIE, CHIRON) are publicly available for this object.

\subsection{Young association}  \label{sec:YA}

\citet{Gagne18} recommend treating candidates with BANYAN $\Sigma$ probability greater than 95\% as probable young stellar association members, so we limit our discussions to the ten candidates with P $>$ 95\% in the following text. TOI 4668 is a member of the Open Cluster Alessi 13 from table \ref{tab:OC}, and in the meantime it is included in Table \ref{tab:YA} as a member of the $X^1$ For (XFOR) moving group. Alessi 13 and the XFOR moving group is synonymous according to Simbad (\citealt{Wenger00}). Similar to Figure \ref{fig:transit_OC},  in Figure \ref{fig:transit_YA} we show phase-folded transit light curves for YA candidates with P $>$ 95\% except for TOI 4668 to avoid redundance. In Figure \ref{fig:transit_YA}, TOI 200, TOI 2221, and TOI 1098 show variations during the transit phase, which are due to the light curves themselves and our data reduction processes. After detrending and normalizing with the \texttt{lightkurve} package, light curves of the three candidates still show large stellar variations. To further reduce the variations and in the meantime avoid erasing the transit signal, we mask the transit and fit to the rest of unmasked data points, so later when we add back the masked transits they still show variations.

\begin{figure*}
	\centering
	\includegraphics[width=0.9\textwidth, trim={8cm 0 8cm 0}]{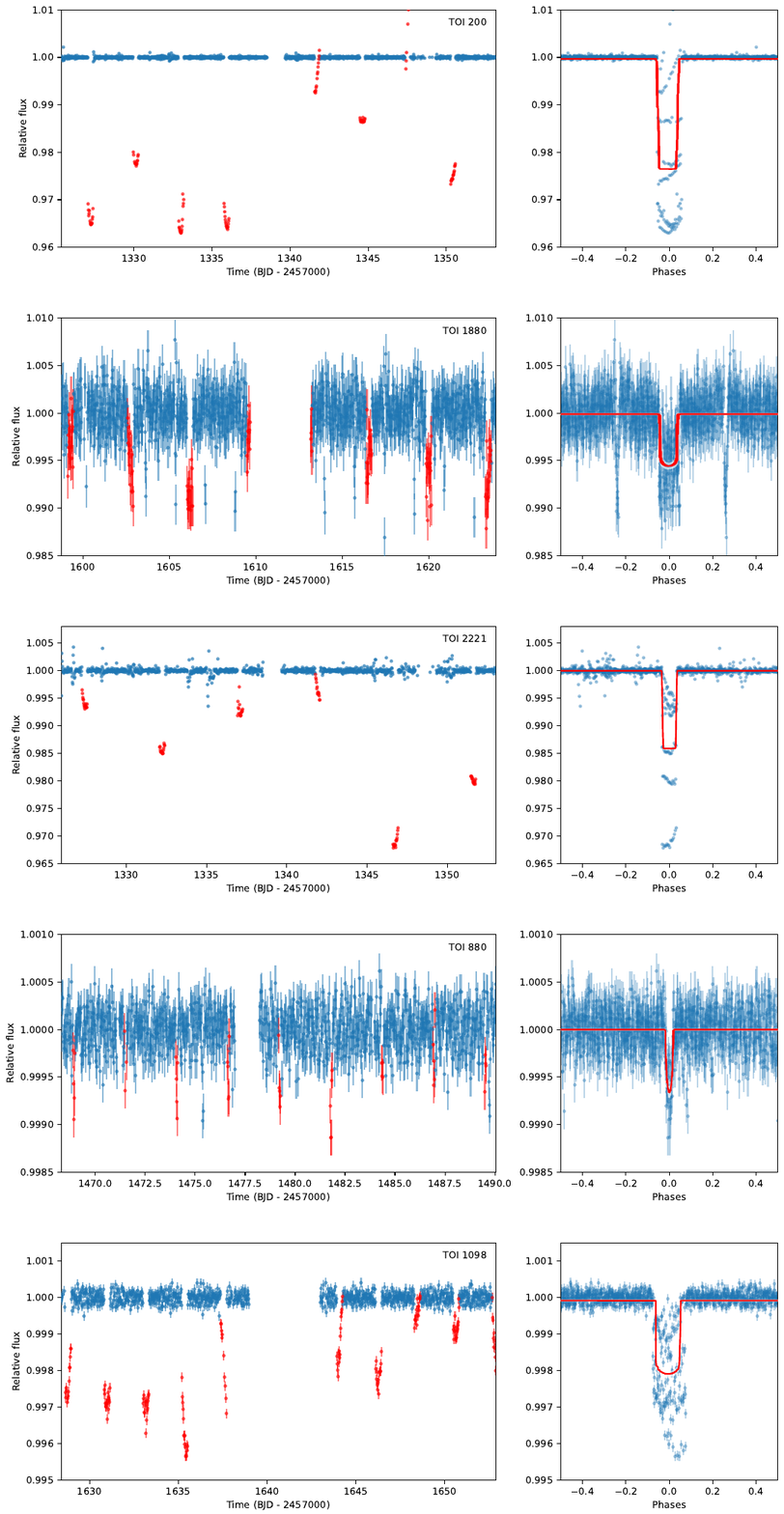}
	\caption{TESS detrended lightcurves (left panel) and phase-folded transit light curves (right panel) for planet candidates in YA. From top to bottom are TOI 200, TOI 1880, TOI 2221, TOI 880, TOI 1098, TOI 1227, TOI 1779, TOI 2427, and TOI 712. The same symbols are used as in Figure \ref{fig:transit_OC}.}
	\label{fig:transit_YA}
\end{figure*}

\begin{figure*}
	\ContinuedFloat
	\centering
	\includegraphics[width=0.9\textwidth, trim={8cm 0 8cm 0}]{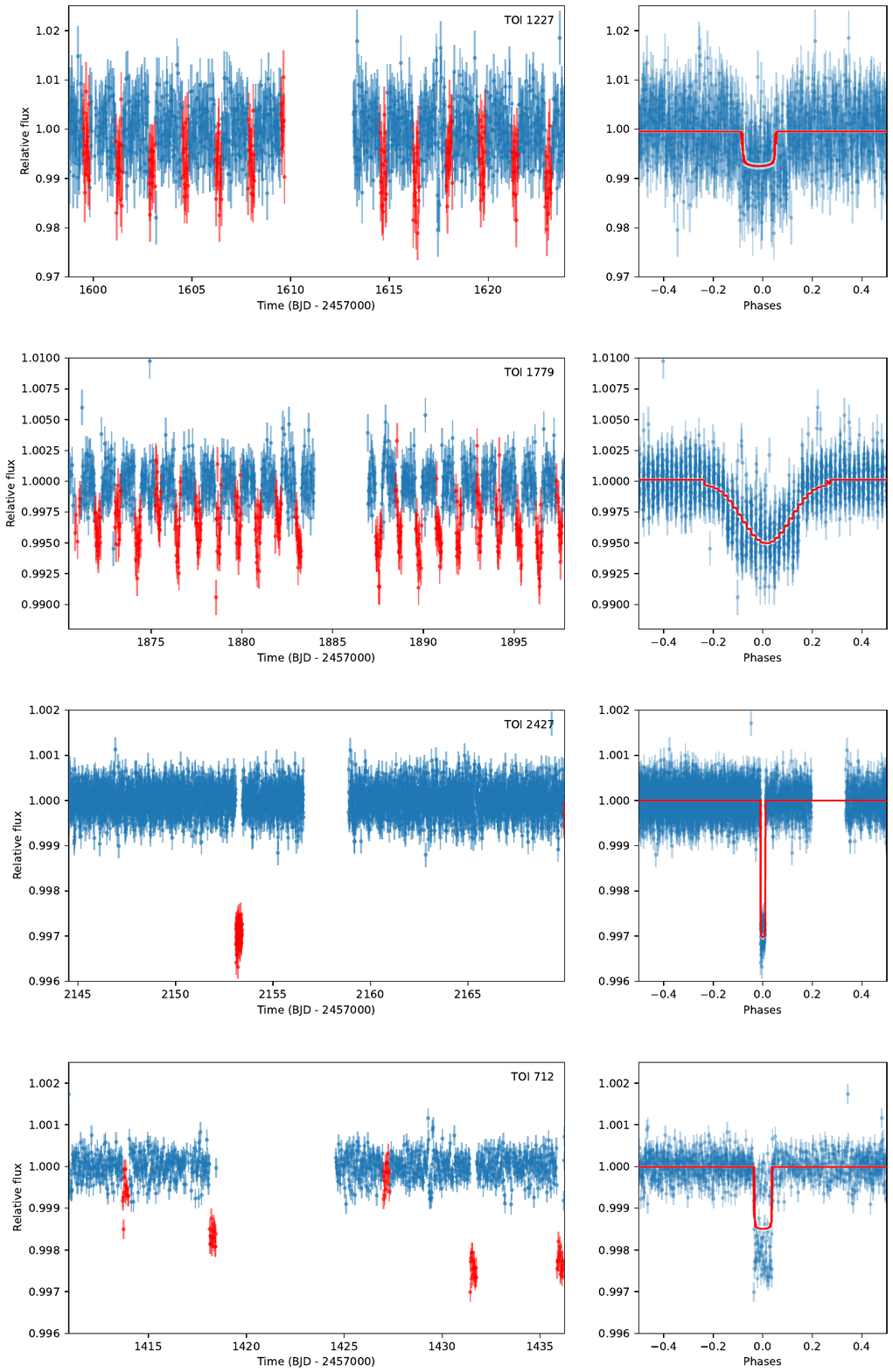}
	\caption[]{TESS detrended lightcurves (left panel) and phase-folded transit light curves (right panel) for planet candidates in YA. From top to bottom are TOI 200, TOI 1880, TOI 2221, TOI 880, TOI 1098, TOI 1227, TOI 1779, TOI 2427, and TOI 712. The same symbols are used as in Figure \ref{fig:transit_OC} (continued).}
	%\label{fig:transit_YA}
\end{figure*}

Five of the candidates host known planets: TOI 200.01 (DS Tuc Ab) in the 45 Myr Tucana–Horologium (THA) young stellar association is discovered by the THYME group (\citealt{Newton19}); TOI 2221.01 (AU Mic b) is a transiting 22 Myr-old Planet orbiting star AU Microscopii in the $\beta$ Pictoris moving group ($\beta$ PMG, \citealt{Plavchan20}); TOI 1227.01 is recently validated by the THYME group to be a member of the Lower Centaurus Crux (LCC) OB association, rather than a member of $\epsilon$ Chamaeleontis (EPSC) from BANYAN $\Sigma$; TOI 1098.01 is a sub-Neptune-sized planet orbiting the young star HD 110082 in the 250 Myr association MELANGE-1 (\citealt{Tofflemire21}); TOI 1779.01 (LP 261-75 b) is a known planet orbiting a low mass brown dwarf (\citealt{RW06}), but its membership to AB Doradus (ABDMG) has not been verified yet. It is promising that the remaining five candidates are young planets belonging to young associations. Table \ref{tab:P_age} lists the proper motion in RA ($\mu_{RA}$), DEC ($\mu_{DEC}$), and parallax for the five candidates.

\begin{table}
	\center
	\begin{threeparttable}
		\caption[]{Period and age for planet candidates in young associations.}\label{tab:P_age}
		\tiny
		\begin{tabular}{cccccccc}
			\hline\noalign{\smallskip}	
			TOI Id & $\mu_{RA}^1$ & $\mu_{DEC}^1$ & parallax$^1$ & (B-V)$_0^1$ & period $^2$ & age $^2$ & YA age$^3$ \\
			& mas yr$^{-1}$ & mas yr$^{-1}$ & mas & mag & day & Myr & Myr  \\
			\hline\noalign{\smallskip}
			TOI 4668 & 35.8232 & -4.51385 & 9.38875 & --$^4$ & -- & -- & -- \\
			TOI 1880 & -39.4012 & -7.24975 & 9.71749 & 1.689 & 3.6 $\pm$ 0.2 & 15.8 $\pm$ 1.7 & 3.7$^{+4.6}_{-1.4}$ \\
			TOI 880 & 0.962399 & 23.2374 & 16.4543 & 0.98 & non-periodic & -- & -- \\
			TOI 2427 & 150.453 & 41.3865 & 35.0407 & 1.393 & non-periodic? & -- & -- \\
			TOI 712 & -2.92948 & 31.0032 & 17.0297 & 1.265 & 5.1 $\pm$ 0.2 & 48.5 $\pm$ 3.7 & 149$^{+51}_{-19}$ \\		
			\noalign{\smallskip}\hline
		\end{tabular}
		\begin{tablenotes}
			\item[1] The proper motion in RA ($\mu_{RA}$), proper motion in DEC ($\mu_{DEC}$), parallax, and (B-V)$_0$ are from the TESS Input Catalog (TIC) queried through MAST.
			\item[2] Rotation period of the host star. Typical error estimation of the period from ACF is $\pm$ 0.2 days, and then the error is propagated to age using equation \ref{eqn1}.
			\item[3] The age of young associations (YA) from \citet{Gagne18}, shown for periodic candidates.
			\item[4] (B-V)$_0$ is not available for TOI 4668.
		\end{tablenotes}
	\end{threeparttable}
\end{table}

\begin{figure*}
	\centering
	\includegraphics[width=1.0\textwidth]{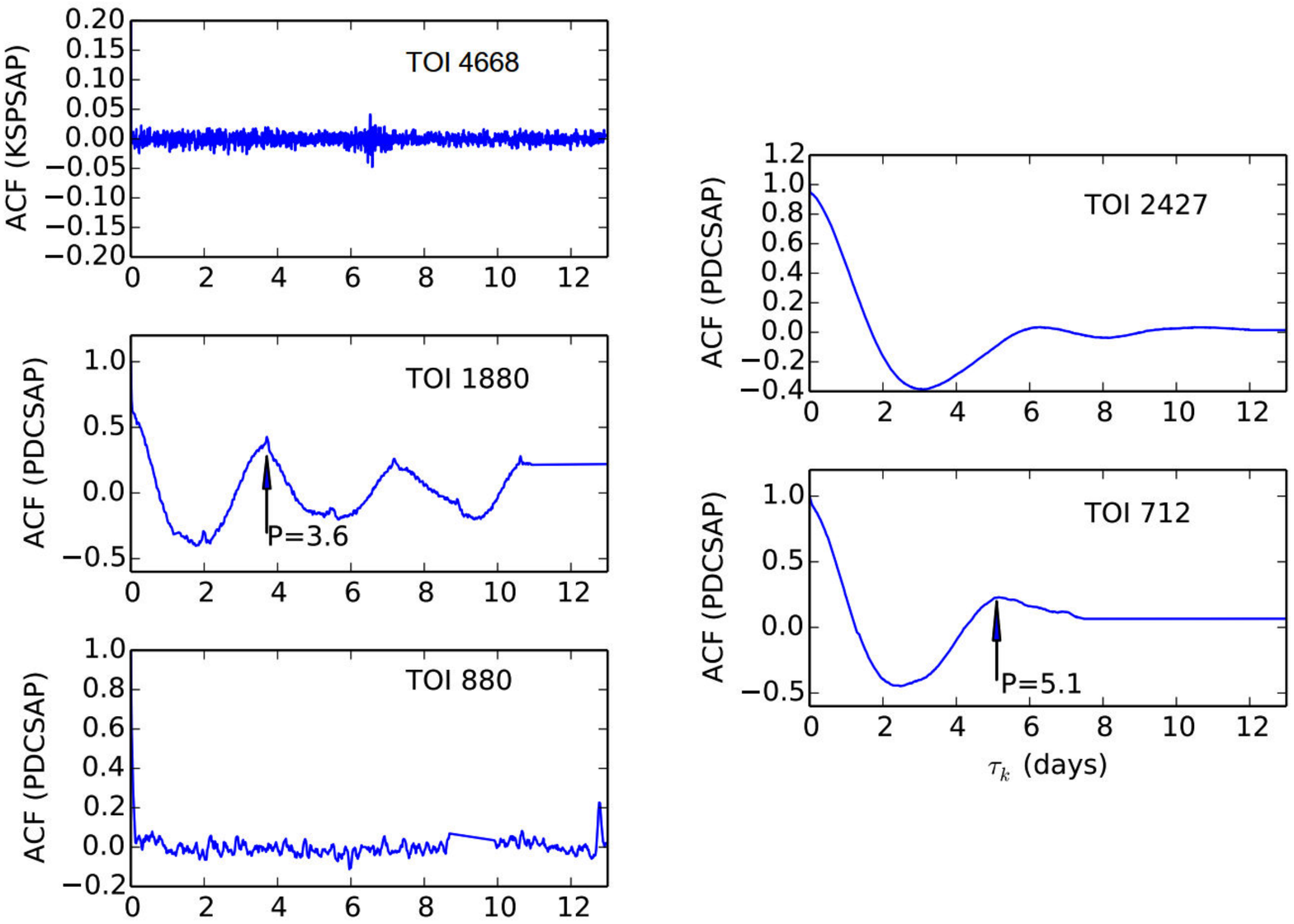}
	\caption{Autocorrelation function for potential candidates in young associations. The TOI Id's of the candidates are marked on each subplot, and the rotation periods for TOI 1880 and TOI 712 are marked on the corresponding subplots.}
	\label{fig:lk}
\end{figure*}

We estimate the rotation periods of the host stars using the TESS light curves with the \texttt{lightkurve} package and estimate their ages based on gyrochronology. We use the autocorrelation function (ACF) method described in \citet{McQuillan14} to roughly estimate the rotation periods for the five planet-hosting candidates in Table \ref{tab:YA} (TOI 4668, TOI 1880, TOI 880, TOI 2427, TOI 712). We check both the Simple Aperture Photometry flux (SAP) and the PDCSAP flux (KSPSAP flux for TOI 4668). The ACFs (KSPSAP flux for TOI 4668, PDCSAP flux for the other five candidates) for each candidate are shown in Figure \ref{fig:lk}.

TOI 4668 and TOI 880 do not show period rotation signals. TOI 1880 shows short period signals in the light curves and ACFs. TOI 2427 is less obvious of a periodic signal from the ACF. Based on its light curve, TOI 2427 seems to only have one transit. TOI 2427 is probably non-periodic, or may have a very long rotation period. TOI 712 is probably periodic. We show the rotation period for TOI 1880 and TOI 712 in Figure \ref{fig:lk} and Table \ref{tab:P_age}.

We then apply the period and Equation \ref{eqn1} (\citealt{Barnes07}; \citealt{RG15}) to estimate the age:

\begin{equation}
log\ t = \frac{1}{n}[log\ P - log\ a - b\ log\ X]
\label{eqn1}
\end{equation}

where P is the period in days, t is age in Myr. We use the gyrochronology relation from \citet{Meibom09} with X = (B-V)$_0$ - c, a = 0.7700, b = 0.553, c = 0.472, and n = 0.5200. The equation applies to (B-V)$_0 \geq$ 0.50 mag and Period $>$ 1.5 days only. We calculate ages for the two candidates showing short period rotation signal, and show the values in Table \ref{tab:P_age}. Ages of the young associations from \citet{Gagne18} are also shown in Table \ref{tab:P_age}.

The ages for both of the two periodic candidates are discrepant with the ages of young associations ($>$ 2$\sigma$). The BANYAN $\Sigma$ Bayesian Algorithm calculates probability based on coordinates, proper motion, and parallax of the planet-star system, which means the candidates of high probability may coincidently satisfy the membership criteria to young associations. The proper motion and parallax might just happen to be close to the young association’s, but the coordinates are actually far away, or vice versa. The three candidates are likely young planets with unknown membership to young associations. 

If the remaining three non-periodic candidates (TOI 4668, TOI 880, TOI 2427) are true young association members, then with their young ages and spectral types we expect to see a clear short period rotation signal in the TESS light curve, but we do not see them. So perhaps these three candidates are not young stars and the young association dispositions are wrong like the other two. In summary, none of the five unpublished systems with P$>$0.95 seems to be really associated with the YA reported by BANYAN $\Sigma$. For systems with P $<$ 0.95, we caution the readers that YA reported by BANYAN $\Sigma$ are unreliable.

\subsection{Number of TOIs in OC and YA}

The planet occurrence rate in OCs/YAs and whether it differs from the occurrence rate in field stars is an important scientific question, which helps understand factors that impact planet formation and evolution in clustering environments (\citealt{Dai21}).  Whether the occurrence rates are different is still under debate due to the small number of planets confirmed in coeval populations, which can lead to large statistical uncertainties.  Using our TOI and OC/YA sample, we would like to give a thought of this question. How many planets do we expect to find in OCs and YAs, and does our result comply with the expections? In the following text of this section we have a rough estimation of the number of planets we expect to find in OCs and YAs. 

First, we assume the ratio of the total number of young stars in OCs and YAs(N$_{young}$) to the number of all TESS stars (N$_{TESS}$) equals the ratio of the number of young TOIs (N$_{TOI, young}$) to the number of all TOIs (N$_{TOI}$), as shown in Equation \ref{eqn:ratio}.

\begin{equation}
	\frac{N_{young}}{N_{TESS}} = \frac{N_{TOI, young}}{N_{TOI}}
	\label{eqn:ratio}
\end{equation}

The number of stars in the TESS Input Catalog is roughly $\sim$ 1.7 billion from the MAST portal (\url{https://mast.stsci.edu/portal/Mashup/Clients/Mast/Portal.html}), after applying a magnitude cut of T = 14 mag, we find 29,911,379 records (data retrieved on April 25 2022). To derive N$_{young}$, we start with a concatenated list of young stars from \citet{Bouma22}, which in their Table 3 includes 1,530,726 young stars. After we apply a similar magnitude cut of G = 14.43 mag by adopting the relationship from equation (2) of \citet[][Tmag = Gmag - 0.430]{Stassun19}, we have 388,207 remaining young stars. We then reasonably assume that 50\% of the young stars are in OC or YA environment, to arrive at N$_{young}$ of 194,104. 

The number of TOIs is 5038, after substituting these numbers into Equation \ref{eqn:ratio}, we find $N_{TOI, young}\ \sim$ 33. This estimation is generally consistent with, but in the meantime a bit higher than, our total number of planet candidates in OCs and YAs. Note that the 10 planets discussed in Section  \ref{sec:YA} all have BANYAN $\Sigma$ P $>$ 95\%, given that their BANYAN $\Sigma$ designations are likely wrong, the other candidates with P $<$ 95 \% may belong to another YA undiscovered by BANYAN $\Sigma$. 

\section{Summary}  \label{sec:summary}

The current TOI Catalog includes over 5000 young planet candidates nearby. We search for planets in 1229 Open Clusters based on the TOI catalog. We find one confirmed planet (TOI 837.01), one brown dwarf (TOI 681.01),one promising candidate (VPC+, TOI 1881.01), and three more unverified due to lack of data (TOI 4668.01, TOI 2538.01, TOI 581.01). For promising candidates, follow-up studies are possible with large telescopes.

We use the BANYAN $\Sigma$ Bayesian algorithm to derive the probability for TOI candidates to be members of nearby young stellar associations. Ten candidates have BANYAN $\Sigma$ membership probability $>$ 95\%, among which five are known planet systems. We use the TESS light curves to derive rotation period and apply gyrochronology method to derive ages for the other five candidates. Two of them show periodic variations likely caused by stellar rotation. Their age discrepancies with the best match young associations suggest these objects are likely young planets with unknown membership to young associations. The other three non-periodic candidates are likely not young stars and the young association dispositions are wrong. The number of planet candidates in OCs and YAs is consistent with our expectation.

\begin{acknowledgements}

We thank an anonymous referee for their constructive comments which helped to improve this work. We thank Elisabeth Newton for her helpful comments on this paper. Q.S. thanks support from the Shuimu Tsinghua Scholar Program. This work is partly supported by the National Science Foundation of China (Grant No. 12133005).

This research has made use of the NASA Exoplanet Archive, which is operated by the California Institute of Technology, under contract with the National Aeronautics and Space Administration under the Exoplanet Exploration Program.

This paper includes data collected by the TESS mission, which are publicly available from the Mikulski Archive for Space Telescopes (MAST). Funding for the TESS mission is provided by NASA’s Science Mission directorate. This research has made use of the Exoplanet Follow-up Observation Program website, which is operated by the California Institute of Technology, under contract with the National Aeronautics and Space Administration under the Exoplanet Exploration Program.

This work has made use of data from the European Space Agency (ESA) mission Gaia, processed by the Gaia Data Processing and Analysis Consortium (DPAC). Funding for the DPAC has been provided by national institutions, in particular the institutions participating in the Gaia Multilateral Agreement.

This research has made use of the VizieR catalogue access tool, CDS, Strasbourg, France. The original description of the VizieR service was published by \citet{Ochsenbein00}. Resources supporting this work were provided by the NASA High-End Computing (HEC) Program through the NASA Advanced Supercomputing (NAS) Division at Ames Research Center for the production of the SPOC data products.

We acknowledge the use of public TOI Release data from pipelines at the TESS Science Office and at the TESS Science Processing Operations Center. We acknowledge the use of public TESS data from pipelines at the TESS Science Office and at the TESS Science Processing Operations Center.

Facilities: Exoplanet Archive, TESS
Software: matplotlib (\citealt{Hunter07}), BANYAN $\Sigma$ (\citealt{Gagne18}).

\end{acknowledgements}

\label{lastpage}

\end{document}